\documentclass[runningheads]{llncs}
\usepackage{graphicx,xcolor}
\usepackage[ruled,vlined]{algorithm2e}
 \usepackage{subcaption}
\usepackage{algpseudocode} % improved pseudo-code
\usepackage{amsmath}
\usepackage{amssymb}
\usepackage{lipsum}
\usepackage{wrapfig}
\usepackage[utf8]{inputenc}
\usepackage{array}
\usepackage{subcaption}
\usepackage{multirow}

\usepackage{afterpage}

\title{\textcolor{red}{\small{This paper has been accepted for publication in Ubiquitous Networking: 7th International Symposium, UNet 2021. Springer International Publishing, 2021. DOI: 10.1007/978-3-030-86356-2\_16}} \\ [2ex] On Computing In the Network: Covid-19 Coughs Detection Case Study}

\author{Soukaina Ouledsidi ali\inst{1} \and
Zakaria Ait hmitti\inst{1} \and
Halima Elbiaze\inst{1} \and
Roch Glitho\inst{2} }

\institute{Université du Québec À Montréal, Montreal, Canada \and Concordia University, Montreal, Canada}

\date{}
\begin{document}
%\authorrunning{Short author list}% Part of LEFT running header
\titlerunning{On Computing In the Network}
\maketitle

\begin{abstract}
Computing in the network (COIN) is a promising technology that allows processing to be carried out within network devices such as switches and network interface cards. Time sensitive application can achieve their quality of service (QoS) target by flexibly
 distributing the caching and computing tasks in the cloud-edge-mist continuum.
This paper highlights the advantages of in-network computing, comparing to edge computing, in terms of latency and traffic filtering. We consider a critical use case related to Covid-19 alert application in an airport setting. Arriving travelers are monitored through cough analysis so that potentially infected cases can be detected and isolated for medical tests.

A performance comparison has been done between an architecture using in-network computing and another one using edge computing.
We show using simulations that in-network computing outperforms edge computing in terms of Round Trip Time (RTT) and traffic filtering.

\end{abstract}


\section{Introduction}
Over the years, the needs of users increase more and more, pushing our global Internet network architecture to its limits. Indeed, low latency, QoS, minimization of core network traffic and scalability have been very challenging to satisfy \cite{NDN}. New approaches have emerged, based on data instead of addresses (IP addresses) such as Information-Centric Networking (ICN). The main idea of ICN is to ensure a communication not through the IP addresses, but rather by sending requests for data without specifying the destination, the network then takes care of retrieve this data in a reliable and secure way, without going each time asking the data to its producer, if the data has already been generated and sent to a closer node, the latter will play the role of the data source \cite{jacobson}.\\
In addition, there is another approach that can be combined with ICN or with the current network protocols, which is in-network computing \cite{noa}, we already know it in the form of load balancer, NAT, caching, and others.
However, it can be more than that, such as processing raw data within network equipment e.g. switches or routers, by doing that, it reduces the consumption of cloud/edge servers’ resources, and also reduces core network traffic. And finally, having as result less congestion and a better latency.
Combining ICN and in-network computing most likely results in a very effective solution.\\
Indeed, an implementation of the combination of these two recent concepts has already been proposed in several research works. Named Function Networking (NFN) \cite{NFN} allows the implementation of one or more functions on any node (as long as it can perform it) of an ICN architecture.
For example, a consumer sends a request specifying that he wants to execute a function with or without parameters, the network then takes the following task: choosing the node that will execute the function, executing it in a reliable and secure way and finally forwarding back the result. All these operations are done by hiding to the end-user all the complexity it takes to make it all work.\\
The purpose of this paper is to implement a use case, that is going to show the advantages of Named Function Networking if implemented in a scenario where there is a low tolerance for high latency.
Our scenario is based on the Covid-19 pandemic situation. Currently, there is a great need to detect infected persons to help reducing the spread of the virus.
One of the gathering points that pose a risk to healthy persons is the airport. Since it represents a place where several persons meet from different countries around the world. Some persons might be carriers of the highly contagious virus inside the airport.\\
Therefore, we designed a use case which makes it possible to detect persons with Covid-19 inside an airport. The use case consists of placing microphones inside the airport corridors to record voices. The recordings are then filtered to extract only coughs for further analysis. Also, next to each microphone there is a surveillance camera placed so whenever an infectious cough is detected the suspicious person will be detected too by analysing the video streaming of the camera.\\
The motivation behind implementing the use case in a NFN architecture is to benefit from a quick processing of audios close to the sources in order to be as efficient as possible to detect infected persons who represent a risk for others. The processing also includes eliminating any speech present in the audios for privacy protection while retaining the cough sounds.\\
The rest of this paper is structured as follows : section 2 presents in-network computing. We present in section 3 the use case overview. Section 4 presents the performance evaluation. In section 5 we present the Discussion. And finally, we conclude the paper in section 6.

\section{In-Network Computing}

\begin{figure}[!h]
   \centering
   \includegraphics[scale = 0.15]{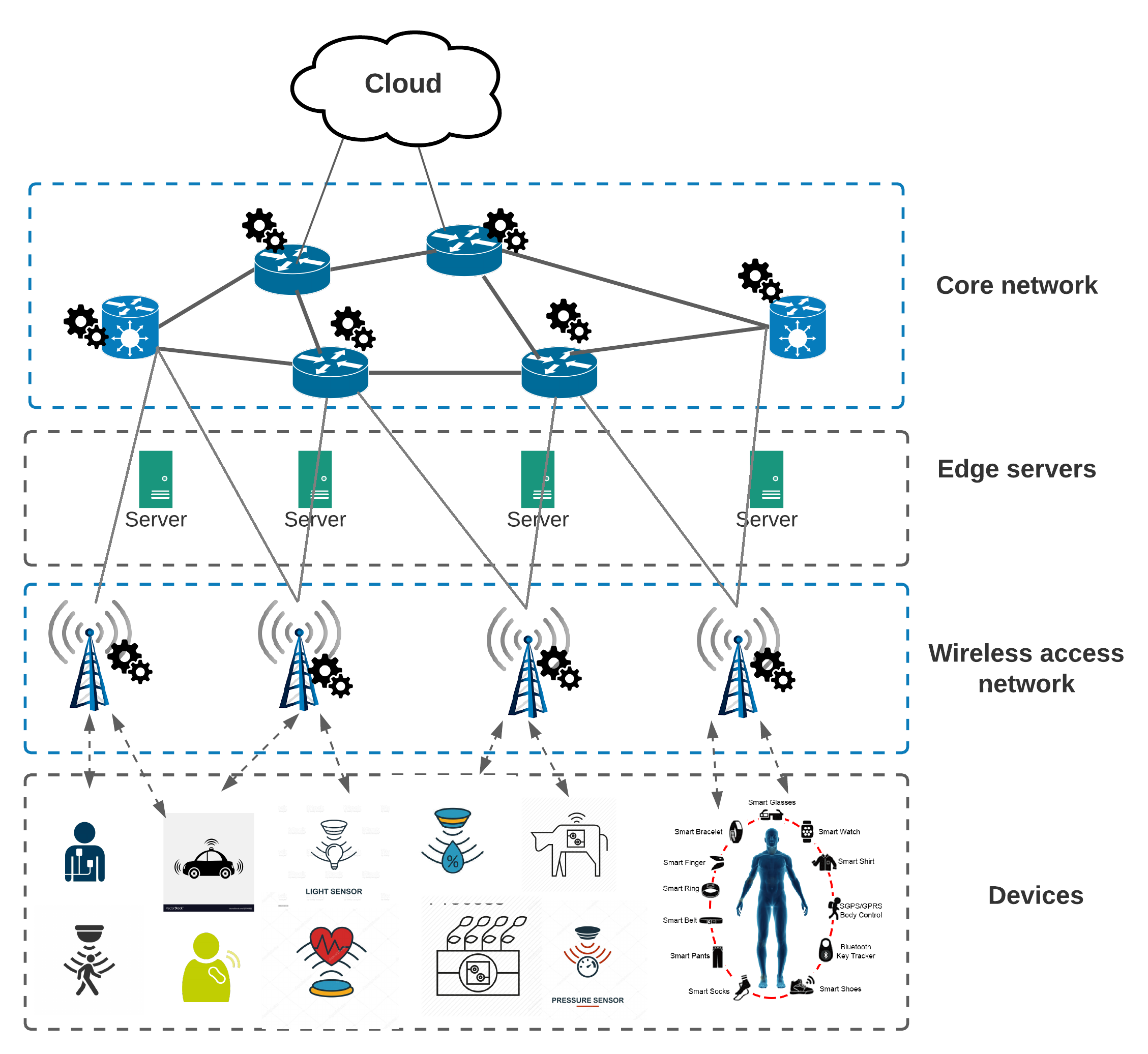}
   \caption{In-network computing architecture}
   \label{System}
\end{figure}

With the evolution of the Internet of Things, cloud computing has started to reach its limits. Objects such as sensors require ultra-fast processing to be able to react in a reasonable time. It is for this reason that edge computing was born. Edge computing pushes computation closer to end users. However, in-network computing which refers to computing within the network, might be more effective in IoT scenarios, for example within mobile network \cite{resolutionstrategies} or critical scenarios that require real-time processing such as AR/VR applications. Figure \ref{System} depicts a 3-tier architecture \cite{NDN}. The tier 1 includes IoT devices e.g. sensors or user's smart devices such as smart phone or smart watch. These devices request data recovery and services. The tier 2 comprises of edge servers responsible for serving end users requests. The tier 3 is the cloud and is located at multiple hops from the tier 1. The cloud has enough resources and high storage capacity but it takes longer to respond to end user requests. In-network computing can be performed in the core network by network devices or in the wireless access network by the access points. The network devices such as routers can process requests that require limited resources and send back the results in a short time.

In the following subsections, we explain our motivation behind the use of in-network computing and we cite works that showed its effectiveness. In the second subsection, we introduce NFN and we explain how it works.

\subsection{Motivation}

Several works have been conducted to create a new extension of the ICN architecture which supports in-network computing, namely NFN \cite{NFN}, Compute First Networking (CFN) \cite{CFN} and Named Function as a Service (NFaaS) \cite{NFaaS}. The aim behind is to benefit from existing mechanisms in ICN, such as caching to cache the results already calculated, which prevents the redundancy of the calculation. Also, we can modify ICN forwarding strategies to do the processing in the best location and have a location-independent execution.\\
As claimed by \cite{ICN-Edge}, in a NFN network, the function code can be cached in the nodes and can move through the network according to the user's request. Also, a requestor can request the execution of a service without having to specify the exact node, which gives flexibility to the network to improve the quality of the service. Several evaluations of NFN have been done in \cite{NFN}, one in a small testbed topology in order to test how NFN distributes function execution tasks, and the benefit of caching results for reuse, as well as showing when is NFN doing network optimization and giving an overhead indication. Experiments have shown the effectiveness of NFN regarding decision making and optimization in different scenarios. \\
NFaaS \cite{NFaaS} concerns the placement and execution of functions using unikernels. It is based on the two paradigms NDN and NFN. In NFaaS, other structures have been added to improve the NFN architecture. These structures are: Kernel store and score function. the Kernel store is responsible for storing functions i.e. unikernels as well as deciding which functions should be executed locally. On the other hand, the score function sort based on a calculated score the functions which are frequently requested in order to know which ones should be downloaded locally.
The scenarios that were made to test the NFaaS framework confirmed the importance of performing functions within the network, especially when it comes to services that are sensitive to delay and bandwidth hungry.\\
Delay-sensitive applications \cite{delaysensitive} are applications that require processing with low latency, such as AR/VR applications, online games or others. In case the device where these applications run has limited resources, their tasks must be offloaded to a near entity such as a network equipment or an edge server.

\subsection{Named Function Networking}

\begin{figure}[!h]
    \centering
    \includegraphics[scale = 0.6]{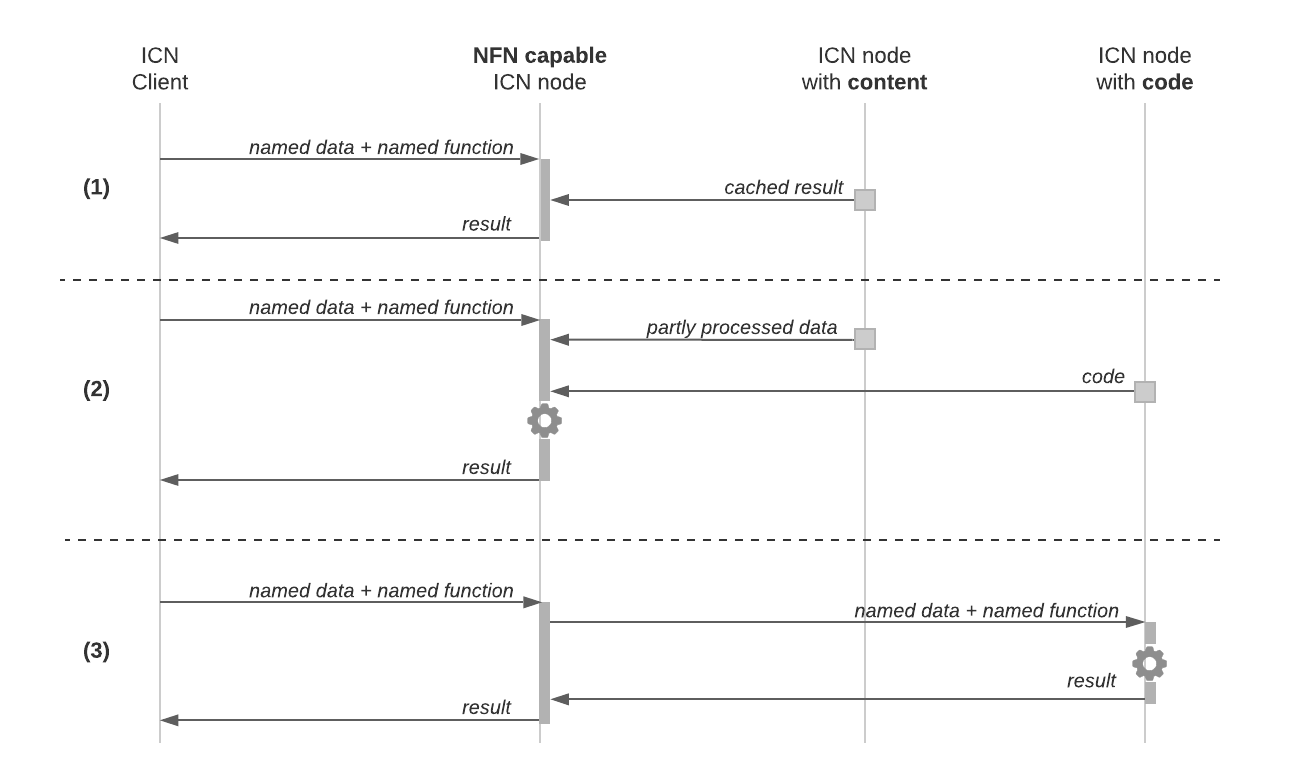}
    \caption{NFN Workflow}
    \label{NFNWorkflow}
\end{figure}

%\textbf{Here you need to talk of how does NFN work
%: explain it in the context of figure 1}\\
NFN allows the orchestration of computation's tasks within the network dynamically, when an ICN client send a request for a computation task three scenarios can happen \cite{NFN} as shown in figure \ref{NFNWorkflow}:\\
(1), the result of the request has already been computed for a previous request, in this case the ICN node retrieve result from content source node.\\
(2), the result doesn't exist or it takes too much time to be retrieved, in this case the node will execute the requested function, and if it doesn't have the code, it will send a request to bring it from a code source node.\\
(3), the request precises a specific site where to address it e.g. due to policy, in this case the network forward the request to the specific site.

%\textbf{You can add a figure that shows the work flow from data %generation to results
%}

\section{Use Case Overview}
The main objective of the use case is detecting persons potentially infected with the corona virus in an airport since it represents a gathering place for different persons from different countries around the world. In this work, we are interested in a specific area of the airport which is the corridors because it is the first area through which the travellers pass after the landing of the plane. These corridors will be equipped with a control and detection system.

\begin{figure}[!h]
    \centering
    \includegraphics[scale = 0.6]{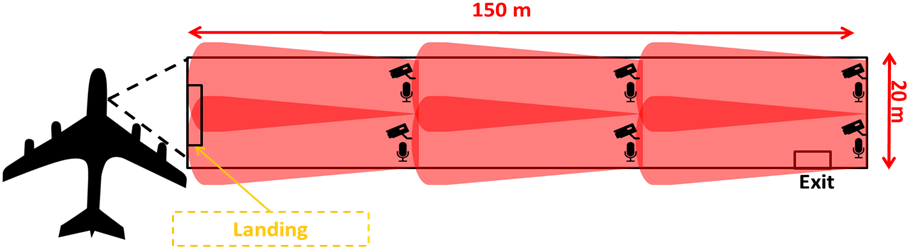}
    \caption{Cameras and microphones placement in the corridor of the airport}
    \label{camerasincorridor}
\end{figure}

We consider Montréal-Pierre Elliott Trudeau International Airport \cite{YUL}. Yul airport has three zones, a zone for domestic flights containing 26 gates, a zone for international flights other than the United States with 12 gates and a zone for flights to the United States containing 18 gates. Each gate is considered to be the opening of a corridor.
Microphones with high sensitivity will be installed along the corridors as shown in figure \ref{camerasincorridor}. Among these types of microphones, we can use the long-range parabolic microphone MEGA EAR \cite{micro} designed by Endoacustica \cite{endoacustica} Europe laboratories which have a range of 50 metres and it can detects the sound of fingers rubbing. For a surface of 150 metres long and 20 metres wide, six microphones will be sufficient to cover the entire area. Near each microphone a camera is placed to record continuously in order to recognize the suspicious person.
All these devices are connected to the edge server via an ICN network.

The sensors i.e. microphones and cameras are connected to the network via access points. The microphone records the voices and send continuously the audio data to the access point which forwards data to the NFN forwarder that is responsible to process these audios. Same operations are done on the side of the camera. 
Using access points provides multiple advantages as explained in \cite{wirelessCommunication} . They can be configured and set up easily, they provide better coverage, more flexible and adaptable compared to wired network, and since they are easy to install and don't require cables, the access points are relatively cost-effective. \\
However, the access points must be placed in an efficient manner in order to benefit from the previous advantages and to support maximum wireless performance. The access points placement and management are outside the scope of this paper.

\subsection{Use case requirements}
%\subsubsection{Round Trip time (RTT)}

\begin{figure}[!h]
    \centering
    \includegraphics[scale = 0.6]{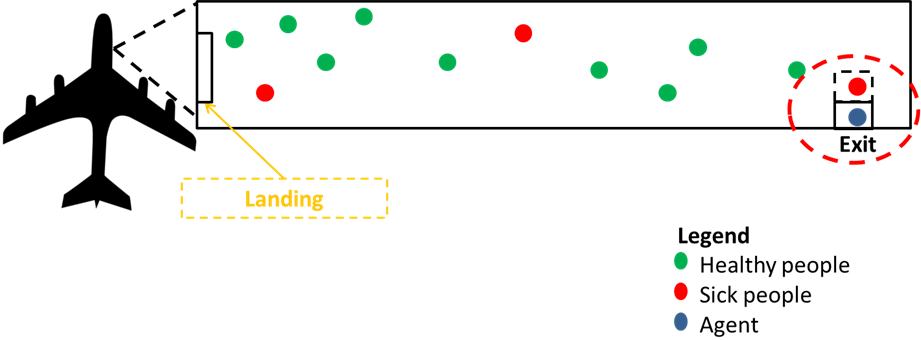}
    \caption{Travelers walking through the corridor}
    \label{peopleincorridor}
\end{figure}

Figure \ref{peopleincorridor} shows travelers walking through the corridor some of them are healthy (green circles), and others are infected with Covid-19 (red circles). The blue circle represents the agent who is going to arrest suspicious travelers.
We assume that if a person leaves the corridor, he will be out of control. So, to deduce the RTT requirement, we decide to study the worst-case scenario which is when a person coughs when he arrives at the agent, in this case the agent should be alerted as soon as possible so that he can stop him before he leaves the corridor. For more details, the protocol of verifying people at the end of the corridor corresponds to: first, the agent checks the traveler's passport, after that he verifies the screen to see if there is any alert considering this person. Then, if this person is suspicious, the agent stops him for further testing. Otherwise, the agent will let him leave the corridor. This interaction takes at least 15 seconds i.e. time needed to proceed in a very fast way. So if the person has an infectious cough just at the moment before the agent started the checking, the system has a maximum of 15 seconds to send the alert to the screen of the agent, before he finishes the checking.

%\subsubsection{Bandwidth}
Furthermore, a recording of a microphone with mp3 format will need a bandwidth of 0.017 Mbytes/s. For the camera, an HD video camera 30 fps that consume at least 0.38 Mbytes/s of the bandwidth will be used, so that the quality will not be degraded.
Considering the corridor shown in figure \ref{camerasincorridor}, for a surface of $3000 m^2$ we deduce that we need a bandwidth of at least 0.4 Mbytes/s for each pair of one microphone and one camera, which means that each corridor of such an area will need 2,4 Mbytes/s.

\subsection{Scenario description}
\begin{figure}[!h]
   \centering
    \includegraphics[scale = 0.35]{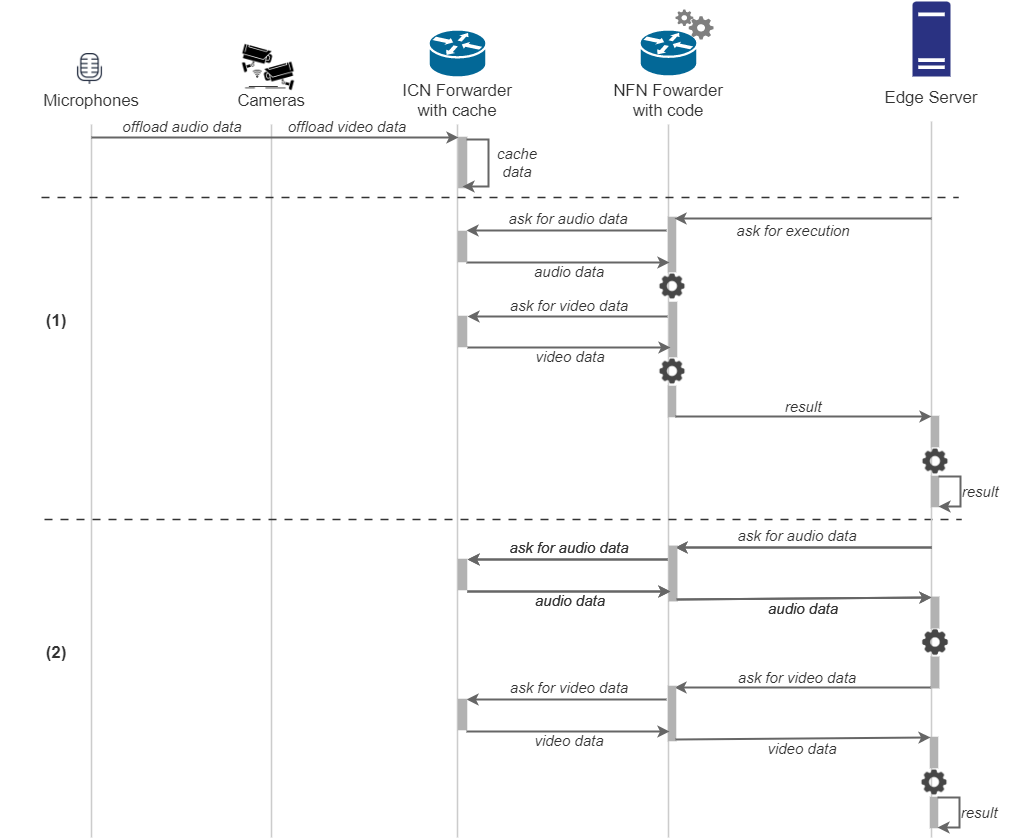}
    \caption{Use case workflow}
    \label{UseCaseWorkflow}
\end{figure}

Let’s assume we have a microphone $M$, a camera $C$, a ICN forwarder $F1$ with caching capability, a NFN forwarder $F2$ contains the code of the cough detection function and capable of executing it and an edge server $E$. $M$ and $C$ stream respectively the sounds and the videos recorded to $F1$.\\
In case we use in-network computing i.e. case (1) in figure \ref{UseCaseWorkflow}: $E$ sends a request asking for the execution of the cough detection function by specifying the microphone ID and the epoch.\\
$F2$ retrieve the audio of that epoch from $F1$ and then analyzes it and decides if it contains an infectious cough or not, by running a deep learning cough detection program. In the case $F2$ detects a cough it automatically sends the video recorded at the time when the cough was detected to $E$. Otherwise, if no cough was detected $F2$ responds with a message indicating that no cough was detected in the specific audio.
If the response is a video, $E$ runs a video analysis program to recognize the suspicious person.\\
In case we use only edge computing i.e. case(2) in figure \ref{UseCaseWorkflow}: $E$ sends a request asking for audio data. Then, it analysis the audio and in case it contains an infectious cough, it asks for video data corresponding to that time where the cough is detected to identify the the suspicious person using a video analysis program.

\section{Performance Evaluation}
%\subsection{Simulation}
Our use case is simulated using PiCN library \cite{picn}. The evaluation is done with two different scenarios: First, using in-network computing i.e. the processing is done by the forwarder, and second, using the edge node for the processing.
\subsection{Settings} 
\begin{figure}[!h]
    \centering
    \includegraphics[scale = 0.3]{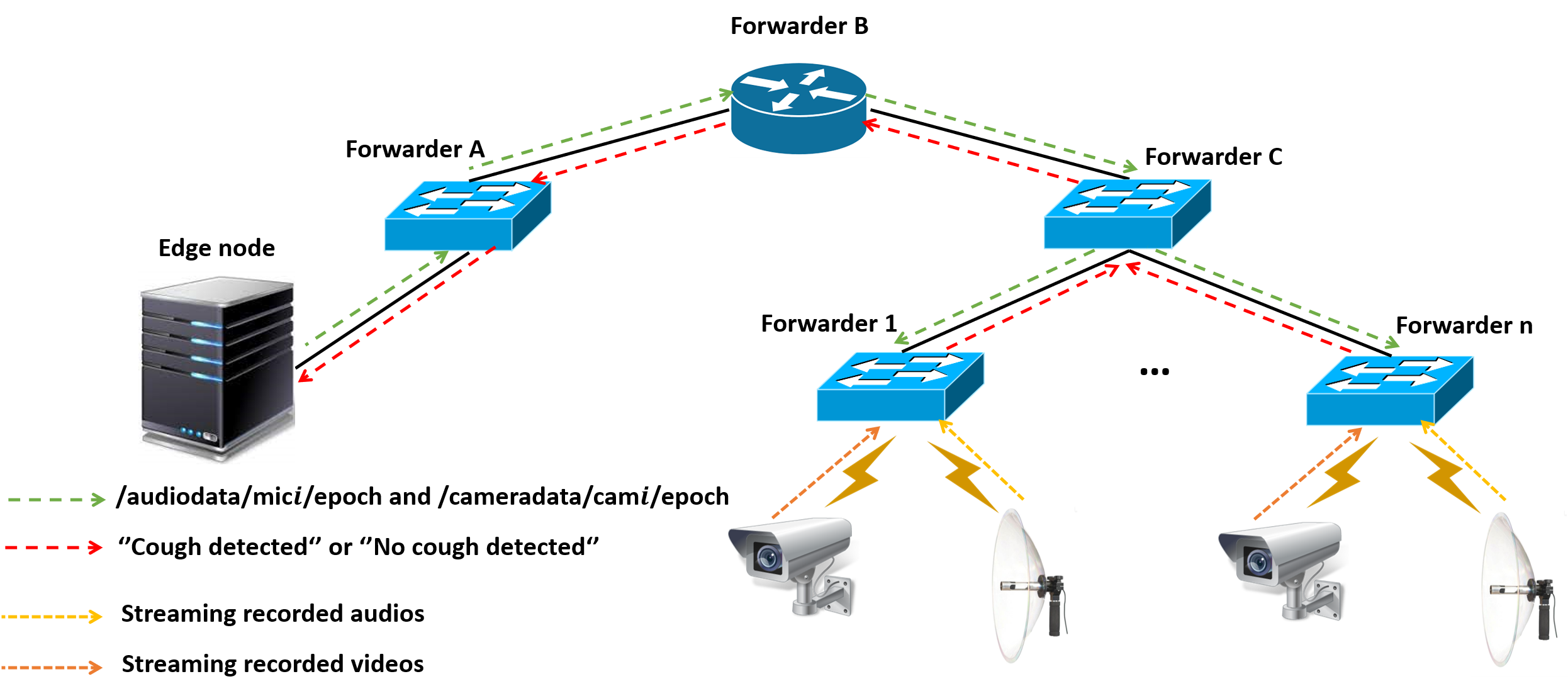}
    \caption{Simulation topology}
    \label{simtop}
\end{figure}

The simulation consists of connecting the microphones and cameras with a forwarder via a wireless network. Each forwarder $j$ with $j \in \{1, \dots, N_c\}$ is a NFN forwarder with $N_c$ is the number of corridors.
%it analyses data coming from the microphones and it sends to the edge node the camera recording of a specific epoch when the cough is detected. 
The forwarder $j$ represents the corridor $j$. As shown in figure \ref{simtop}, all these forwarders are connected to a core switch represented by forwarder C. The forwarders B and A insure packets routing between the corridors' network and the edge node. Additionally, each forwarder caches the paths of audios identified by the epoch of the time where they were recorded and the paths of videos corresponding to those audios. We use a sample of fixed number of audios and videos, each one of them have a duration of 5 seconds. Some audios were recorded with coughs and others with simple talks. The videos consist of recordings of travelers walking throughout an airport. \\
The edge node asks for the execution of the function detectCough() on audio recorded at a specific epoch. The audio is recorded by the microphone $i$ with $i \in \{1, \dots, N_{mic}\}$ and $N_{mic}$ is the number of microphones inside a corridor. The edge node sends an NFN interest to forwarders from 1 to n. The interest follows the path specified in the Forwarding Interest Base (FIB), until it reaches forwarders from 1 to n.\\
In in-network computing scenario, when the forwarders from 1 to n receive the interest message, they check in their Content Store (CS) if they have the function and the content of this naming on the CS: /mic$i$/epoch. Otherwise, the interest is forwarded to the next forwarder.
After that, the forwarder starts the execution of the function which consists of a deep learning program that predicts if the audio contains a cough or not. When the execution is done, the forwarder sends to the edge server the result returned by the function. The result is in the case of "Cough detected", the video recorded at the time when the cough was detected. Otherwise, if "No cough detected", the forwarder responds with a message. The response follows the reverse path used to forward the interest message.

%\subsubsection*{Edge Computing scenario}
On the other hand, in edge computing scenario, the edge node sends interests to the forwarder A asking for the audio path. The forwarder A forwards the interest to the other forwarders.
%Then, each forwarder checks in its CS to see if it has the audio path, if it does not find it, it  forwards the interest to the next forwarder, following the paths in the FIB, and it goes so on until it reaches a forwarder from 1 to n. 
When a forwarder from 1 to n receive the interest message, it checks its CS, and send back the content to the edge server. The edge server will execute the cough detection program and if it finds a cough on it, it will send another interest to recover the video corresponding to the audio when the cough was detected, by sending an interest message to forwarder A. And finally, it brings the video for further analysis.

\subsection{Deep learning based cough detector mechanism}
The cough detection program \cite{deeplearning} consists of two python programs "Training.py" and "Inference.py". First, "Training.py" gets as inputs 162 non-cough sounds i.e. different environmental sounds, that we got from ESC-50 \cite{ESC} dataset which is a publicly available dataset, and 53 cough sounds some of them are from Covid-19 \cite{covid19} dataset. With the cough sounds we have a CSV file which contains for each audio file, the time laps where there is a cough. Then, we train the program and it generates a model. "Inference.py" will be based on this model to predict for each audio recording, if it has a cough or not.
The audios giving as input to "Inference.py" were recorded by us, some of them contain a speech with coughs and others only speech.\\
Notice that the data used for the training might not be the optimal one, but we focused more on the network part of the simulation rather than the cough detection program.

\subsection{Numerical Results}
\subsubsection*{Round trip time (RTT)}
In the simulation we use six pairs of one microphone and one camera, each pair has a sample of 30 audios with cough sound of 5 seconds, and 30 videos each one related to one microphone of the same duration. 
%\\In the simulation program, we monitored each pair with a different thread, and we calculated the RTT for every audio.
\\When a cough is detected, the RTT in case of in-network computing is calculated as the sum of the time needed to request for filtered audio data, run detectCough() function on forwarder near the microphone and response with the video captured in that time.
In case of edge computing, the RTT is the sum of the time needed to request for audio data, run detectCough() function on the edge node, return a result specifying that a cough is detected, request for video captured in that time and recover video data.

\begin{figure}[!h]
    \centering
    \includegraphics[scale = 0.5]{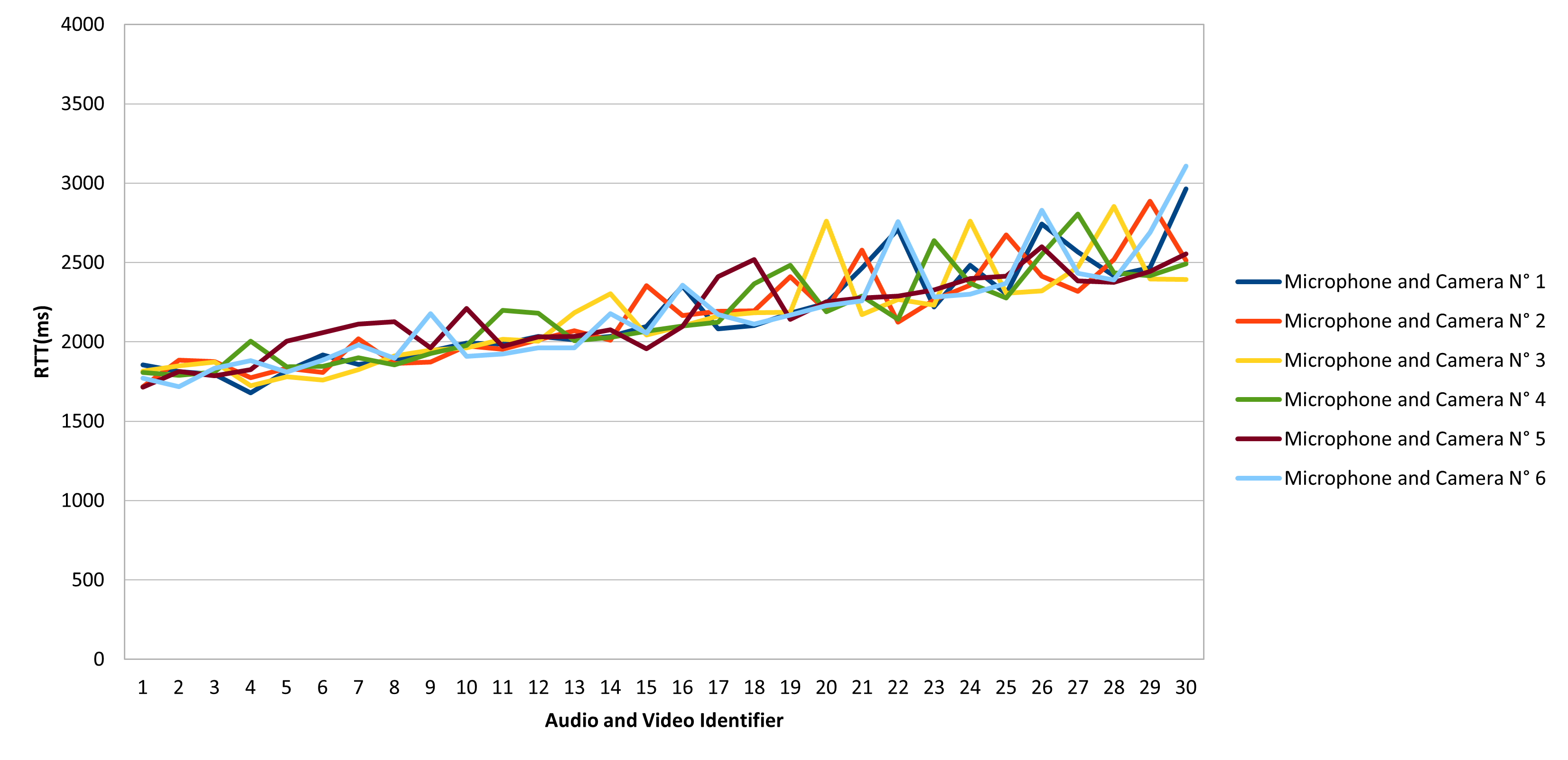}
    \caption{RTT in in-network computing scenario}
    \label{resultsWithInc}
\end{figure}

Figure \ref{resultsWithInc} depicts the RTT in milliseconds for each pair of one microphone and one camera in case of in-network computing. The RTT for all the pairs is increasing approximately the same way and it is mainly due to the processing time of the audios at the forwarder node. The processing is different from one audio to another as each audio has a particular speech with a different cough. The average of RTT for all pairs is less than 2.2s which satisfies the requirements discussed in the sub-section 3.1.

\begin{figure}[!h]
    \centering
    \includegraphics[scale = 0.5]{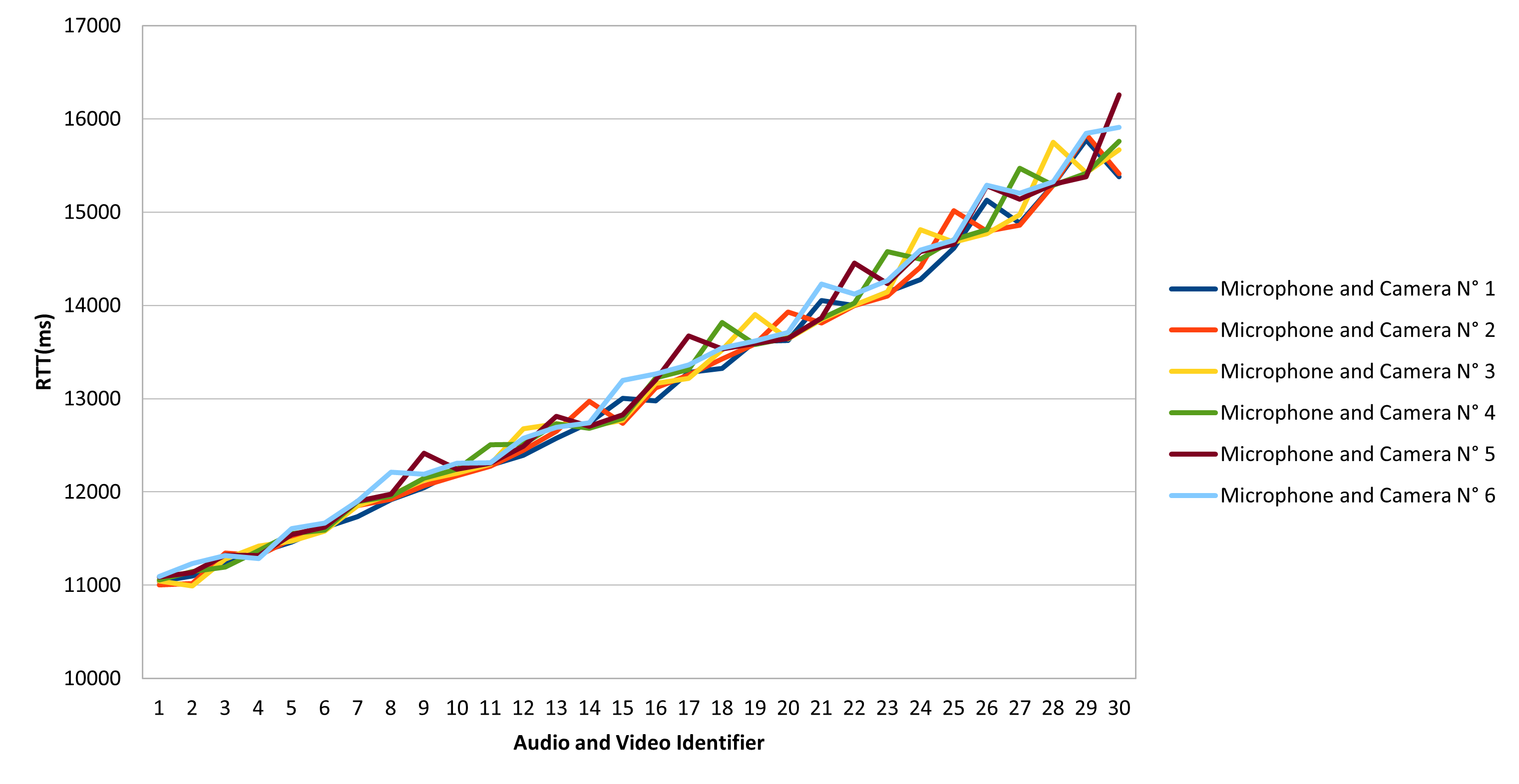}
    \caption{RTT in edge computing scenario}
    \label{resultsWithoutInc}
\end{figure}

Figure \ref{resultsWithoutInc} shows the RTT in milliseconds for each pair of one microphone and one camera in case of edge computing. We can clearly see that for the first audio and video identifier, the RTT is near 11 seconds and starts to exceed the required RTT i.e. 15 seconds from approximately the audio and video identifier 27. The increase of the RTT is very significant when asking for processing to the edge node. This increase is due to the processing of audios and the time taken for both the requests for the audio and the video related to.

\subsubsection*{Traffic filtering and memory consumption}
Referring to sub-section 3.1, we concluded that each corridor equipped with 6 pairs of microphone and camera needs 2,4 Mbytes/s of bandwidth, due to audio and video data streaming. We notice that when we execute our function on the NFN forwarders near to corridors, the forwarder filter audio and video streaming data, and sends only videos of potentially infected people.
Additionally, we did measure the memory consumption of our program executed in parallel on 6 audio/video data streaming input, and it has an average of 2,25 GBytes of memory allocated for the different processes.
Assuming that we have an airport of only $N_c = 10$ corridors, it's almost 22,5 GBytes in terms of memory that can be gained on edge server, if we use in-network computing paradigm.

%\section{Discussion}
%Based on the experiences we have presented, it is easy to realize that in-network computing has several advantages and reacts better in time sensitive scenarios. What mainly makes it advantageous is the physical distance between requesters and the nodes responsible for processing. However, network devices have limited resources, and this can restrict network performance in some particular cases.
%When there is a need to perform a function that is resource intensive, it would be better in this case to delegate the task to the edge servers. Additionally, edge servers have higher storage capacity, so a program that needs a lot of data which are stored in the edge server, it would be more efficient to run it in that node instead of fetching a large batch of data.
%A scalability test could perhaps show that edge computing adapts better with the increase in the number of requests. Since it has enough resources to serve several requests at the same time. On the other hand, an experiment could show that parallel processing carried out by several forwarders at the same time, can decrease the overall network throughput.

\section{Conclusion}
In-network computing is a new research area that brings a lot of promises, especially when applied in Information-Centric Networking architecture. NFN extends ICN to execute functions. In this work, we compare in-network computing with classical edge computing in ICN architecture, by designing a use case that allows to detect persons infected with Covid-19 inside an airport. We have implemented an architecture designed of sensors i.e. cameras and microphones, forwarders and edge server, and a deep learning based program which is executed within the network to detect suspicious coughs. The results show that in-network computing performs better than edge computing as the processing is done in a minimal delay and largely satisfies the use case requirements.

\section*{Acknowledgement}
This work is partially supported by Chist-Era SCORING-2018 Project.

\end{document}